\documentclass{sigchi}

\toappear{\scriptsize Permission to make digital or hard copies of all or part of this work for personal or classroom use is granted without fee provided that copies are not made or distributed for profit or commercial advantage and that copies bear this notice and the full citation on the first page. Copyrights for components of this work owned by others than ACM must be honored. Abstracting with credit is permitted. To copy otherwise, or republish, to post on servers or to redistribute to lists, requires prior specific permission and/or a fee. Request permissions from permissions@acm.org. \\
{\emph{CHI '20, April 25--30, 2020, Honolulu, HI, USA.} } \\
Copyright is held by the owner/author(s). Publication rights licensed to ACM. \\
ACM ISBN 978-1-4503-6708-0/20/04\ ...\$15.00.\\
https://doi.org/10.1145/3313831.3376257
}

\toappear{\scriptsize Permission to make digital or hard copies of all or part of this work for personal or classroom use is granted without fee provided that copies are not made or distributed for profit or commercial advantage and that copies bear this notice and the full citation on the first page. Copyrights for components of this work owned by others than the author(s) must be honored. Abstracting with credit is permitted. To copy otherwise, or republish, to post on servers or to redistribute to lists, requires prior specific permission and/or a fee. Request permissions from permissions@acm.org. \\
{\emph{CHI '20, April 25--30, 2020, Honolulu, HI, USA.} } \\
\copyright~2020 Copyright is held by the owner/author(s). Publication rights licensed to ACM. \\
ACM ISBN 978-1-4503-6708-0/20/04\ ...\$15.00.\\
http://dx.doi.org/10.1145/3313831.3376257 }

\usepackage{balance}       
\usepackage{graphics}      
\usepackage[T1]{fontenc}   
\usepackage{txfonts}
\usepackage{mathptmx}
\usepackage[pdflang={en-US},pdftex]{hyperref}
\usepackage{color}
\usepackage{booktabs}
\usepackage{textcomp}

\usepackage{microtype}        
\usepackage{ccicons}         
\usepackage{todonotes}

\usepackage{multirow,booktabs}

\usepackage{subcaption}
\usepackage{amsmath}
\usepackage{mdframed}
\usepackage{float}
\usepackage{xcolor,colortbl}
\usepackage{threeparttable}
\usepackage{lmodern}
\mdfdefinestyle{MyFrame}{
    innerrightmargin=6pt,
    innerleftmargin=6pt,
    innertopmargin=6pt,
    innerbottommargin=6pt,
    linewidth=1pt
}

\usepackage[format = plain, font=small, labelfont=bf]{caption}
\def\plaintitle{Do I Look Like a Criminal? Examining how Race Presentation Impacts Human Judgement of Recidivism}
\def\plainauthor{First Author, Second Author, Third Author, Fourth Author, Fifth Author, Sixth Author} 

\def\plainkeywords{Authors' choice; of terms; separated; by semicolons; include commas, within terms only; this section is required.}

\makeatletter
\def\url@leostyle{%
  \@ifundefined{selectfont}{
    \def\UrlFont{\sf}
  }{
    \def\UrlFont{\small\bf\ttfamily}
  }}
\makeatother
\def\pprw{8.5in}
\def\pprh{11in}

\definecolor{linkColor}{RGB}{6,125,233}
\hypersetup{%
  pdftitle={\plaintitle},
  pdfauthor={\plainauthor},
  pdfkeywords={\plainkeywords},
  pdfdisplaydoctitle=true, 
  bookmarksnumbered,
  pdfstartview={FitH},
  colorlinks,
  citecolor=black,
  filecolor=black,
  linkcolor=black,
  urlcolor=linkColor,
  breaklinks=true,
  hypertexnames=false
}

\newcommand\emptyfootnote[1]{%
  \begingroup
  \renewcommand\thefootnote{}\footnote{#1}%
  \addtocounter{footnote}{-1}%
  \endgroup
}

\begin{document}

\title{\plaintitle}
\numberofauthors{6}
\author{%
  \alignauthor{\stepcounter{footnote}Keri Mallari\thanks{\scriptsize Work done while interning at Microsoft Research.\bigskip}\\
    \affaddr{University of Washington}\\
    \email{kmallari@uw.edu}}\\
  \alignauthor{Kori Inkpen\\
    \affaddr{Microsoft Research}\\
    \email{kori@microsoft.com}}\\
  \alignauthor{Paul Johns\\
    \affaddr{Microsoft Research}\\
    \email{pauljoh@microsoft.com}}\\
  \alignauthor{Sarah Tan\\
    \affaddr{Cornell University}\\
    \email{ht395@cornell.edu}}\\
  \alignauthor{Divya Ramesh$^\dagger$\\
    \affaddr{University of Michigan}\\
    \email{dramesh@umich.edu}}\\
  \alignauthor{Ece Kamar\\
    \affaddr{Microsoft Research}\\
    \email{eckamar@microsoft.com}}\\
}
\maketitle

\begin{abstract}
\emptyfootnote{}Understanding how racial information impacts human decision making in online systems is critical in today's world. Prior work revealed that race information of criminal defendants, when presented as a text field, had no significant impact on users' judgements of recidivism~\cite{dressel2018accuracy}. We replicated and extended this work to explore how and when race information influences users' judgements, with respect to the saliency of presentation. Our results showed that adding photos to the race labels had a significant impact on recidivism predictions for users who identified as female, but not for those who identified as male. The race of the defendant also impacted these results, with black defendants being \textbf{less} likely to be predicted to recidivate compared to white defendants. These results have strong implications for how system-designers choose to display race information, and cautions researchers to be aware of gender and race effects when using Amazon Mechanical Turk workers.
\end{abstract}


\section{Introduction}
In today's online society, people are frequently asked to make judgements about others based on online profiles \cite{donath2007signals, fagerstrom2017personal, bakhshi2014faces}. Whether it is rating Uber drivers, selecting applicants to interview for a job, or deciding whether someone gets approved or rejected for a loan, people are frequently asked to evaluate others. 

In 2018, Dressel and Farid \cite{dressel2018accuracy} compared human accuracy to the accuracy of a commercial AI system for recidivism prediction. They showed Amazon Mechanical Turk workers (whom we refer to as users in the rest of this paper) real defendant profiles and asked them to rate whether or not they thought that a defendant would commit a crime in the next two years (i.e. recidivate). Two different variants of the study were run, one where users were told the race of the defendant, and one where they were not. The authors found no significant differences in the results of these two variants and concluded that including race information had no impact on the accuracy or fairness of the users' decisions. 

Concluding that defendant race has no significant effect on human judgements of recidivism is risky given the complexity of racial bias, especially on issues related to criminal justice~\cite{skeem2016risk, kleider2017black, schwalbe2006classifying, smith1984equity, rachlinski2008does}. However, Dressel and Farid's result is curious.  Is it the case that race doesn't matter for predictions of recidivism? Probably not~\cite{wehrman2010race}. Is it the case that the users in Dressel and Farid's study were not biased? Probably not~\cite{wilson2017racial}. 
Is it the case that the Amazon Mechanical Turk workers were not engaged in the task such that race did not matter?  Possibly, but given how common implicit biases are \cite{teachman2001implicit,greenwald2006implicit,kang2011implicit}, it is unlikely that race would have no effect. These questions motivated us to investigate the validity and generalizability of Dressel and Farid's results and the implications it has on the design of evaluation systems, as well as studies utilizing Mechanical Turk workers. 

Improving our understanding of biases related to human judgement in online systems is important given that the use of AI systems to assist human decision-making is growing, especially for critical domains such as medicine \cite{cai2019hello}, criminal justice \cite{eckhouse2019layers, green2019disparate}, and social welfare \cite{chouldechova2018casestudy}. Informed design of such systems and developing insights about potential biases they may introduce on decision-making hinges on advancing our understanding of how sensitive attributes affect both sides of Human-AI systems.

In this paper we present three studies. First, a replication of Dressel and Farid's study \cite{dressel2018accuracy} to confirm their experimental protocol and to validate the replicability of their results. Next, we describe Study 1, an extension of Dressel and Farid's work to examine the impact of adding photos to defendant profiles (which serves as a way to increase the prominence of racial cues). Finally, we describe Study 2, which isolated the impact of race from other defendant characteristics by controlling the race of defendant profiles. The key takeaways are:

\begin{itemize}
    \item Showing race information had a significant impact on users' predictions of whether a defendant would recidivate. When both a photo and a race label were added to a defendant's profile, users were \emph{less} likely to predict that they would recidivate [Study 1].
    
    \item Users' self-identified gender was a significant factor in recidivism predictions, with females being impacted when a photo and race label were provided but not males [Study 1 \& 2].  
    
    \item The race of the defendant also impacted female users' recidivism predictions, with black defendants being rated as \emph{less} likely to recidivate compared to white defendants [Study 1 \& 2]. 

    \item The two variants of our No Race condition produced different effects. In Study 1, when users were only shown text, defendants were rated higher in terms of recidivism than when a photo and race label were provided. In study 2, when an ``Image Unavailable'' icon was shown in the no race condition, and the trials were intermixed with other trials that included photos and race labels, recidivism predictions were lower. 
    
\end{itemize}

\section{Related Work}
A large body of work in HCI argues for careful considerations of issues of bias in the design of technology~\cite{friedman1996bias, kilbourne1997socio, hankerson2016does}. Biases, implicit or explicit, frequently limit opportunities for certain groups~\cite{zenk2005neighborhood, martin2006bias}. Developing solutions to mitigate bias is currently a major challenge in  AI~\cite{caliskan2017semantics, propublica, schlesinger2018let, olteanu2019social, buolamwini2018gender}. Implicit gender and racial biases in technology is being studied by both AI and HCI researchers~\cite{mitchell2019model, schlesinger2018let, olteanu2019social, jahanbakhsh2020bias, peng2019you}, and especially in the context of recidivism following ProPublica's report that COMPAS, a nationwide criminal risk assessment tool, contained racial disparities in its predictions~\cite{propublica, van2019crowdsourcing, green2019disparate, dodge2019explaining, chouldechova2017fair, tan2018distill, tan2018investigating}.

\subsection{Risk-Assessments and Judgement on Mechanical Turk}
Recent studies on biases in human and algorithmic risk-assessment have utilized Amazon Mechanical Turk~\cite{dressel2018accuracy, van2019crowdsourcing, green2019disparate} as participants. However, such experiments have yielded mixed support for the claim of racial disparities in user judgements. For instance, Dressel and Farid, when comparing human and algorithmic accuracy for recidivism predictions, did not find race to be a factor influencing human judgement~\cite{dressel2018accuracy}, while Green and Chen~\cite{green2019disparate} found that users were more likely to predict higher recidivism risk for one racial group (i.e. black defendants), when aided with a risk assessment tool. Algorithmic aids have also been shown to bias workers' predictions of recidivism~\cite{grcic2019human,vaccaro2019effects}.

Our work informs the design aspects of such systems and studies, in the context of human and algorithm assessments of recidivism.

\subsection{Impact of Information Presentation on Human Judgement}

Prior research in human decision making has shown that the modality in which information is presented can affect user performance and judgement~\cite{cao2009modality, hartmann2008framing, wobbrock2019isolating, barrett2005picture, kay2015unequal, hong2004designing}. 
For instance, images were found to be more effective than text in presenting emergency and disaster information to users~\cite{cao2009modality}; user perceptions of website quality were found to be influenced by subtle manipulations such as framing~\cite{hartmann2008framing}; and perceived credibility of news websites were found to be affected by purely presentational factors~\cite{wobbrock2019isolating}. 

Similar findings have been discovered about human judgement in the offline world. Newspaper photographs of political candidates can shape voters' perceptions of candidates and alter their likelihood of voting for them~\cite{barrett2005picture}. People’s perceptions about gender distributions in careers in the real-world can shift by modifying representations of gender in career-related image search results online~\cite{kay2015unequal}.

Combinations of modalities and cognitive load also influence user judgements. Image + text information can cause information overload and distractions under high cognitive loads~\cite{cao2009modality}, but improve recall of information under low cognitive loads~\cite{hong2004designing}. Expanding this line of literature, our work explores how the presentation of racial information influences human judgements of recidivism. 

\subsection{Gender and Race Interaction in Human Judgements}
Racial disparities have been studied in domains such as  health, education, crime rates and economic status~\cite{wehrman2010race, williams1997race}. However, race is also a carrier of historical and cultural biases~\cite{kleider2012looking, dixon2005skin}. Kleider et al. found associations between facial features and people's perception of criminal stereotypes~\cite{kleider2012looking,kleider2017black}. In addition, biases can have different effects across gender. In Madriz's study, women reported higher levels of fear of crime than men, and their perceptions of criminality and victimhood were affected by their own social position and race conveyed in images~\cite{madriz1997images}. Similarly, Barrett et al. found that women were more likely to be influenced by positive pictures and men by negative ones when assessing political candidates ~\cite{barrett2005picture}. Our work adds to this literature by exploring the influence of race and gender on human judgements of recidivism. 

\section{Replication Study}
Our goal was to investigate how the presentation of race information impacts human judgement of recidivism. To do so, we extended Dressel and Farid's study \cite{dressel2018accuracy} by presenting race information in  different forms to the user. Before extending Dressel and Farid's study, we first ran a replication study to confirm that our experimental protocol was consistent, and that we were able to replicate their results.

\subsection{Dressel \& Farid Experimental Protocol}
\label{sec:dartmouthprotocol}
Below is a brief summary of Dressel and Farid's experimental protocol. See their paper \cite{dressel2018accuracy} for further details.

Dressel and Farid selected a subset of 1,000 defendants from the ProPublica COMPAS dataset of 7,214 pretrial defendants from Broward County, Florida \cite{propublica}. These defendants were grouped into twenty groups of fifty defendants each. Twenty users were assigned to each group, for a total of four hundred users. Each user evaluated all fifty defendants in their assigned group.

Defendant profiles were presented to users in a descriptive paragraph, as shown below, comprising eight features as well as a description of the crime the defendant was charged with. Dressel and Farid ran two versions of this experiment, one that included defendant race information, and another without. Note that the race label is colored \textcolor{blue}{blue} below for demonstration purposes, but was not blue in the actual presentation.

\begin{center}
    \noindent
    \fbox{
        \begin{minipage}[t]{0.45\textwidth}
        The defendant is a \textcolor{blue}{[RACE]} [SEX] aged [AGE]. They have been charged with: [CRIME CHARGE]. This crime is classified as a [CHARGE DEGREE]. They have been convicted of [NON-JUVENILE PRIOR COUNT] prior crimes. They have [JUVENILE-FELONY COUNT] juvenile felony charges and [JUVENILE-MISDEMEANOR COUNT] juvenile misdemeanor charges on their record.
    
        \vspace{0.5cm}
        [CRIME CHARGE] : [CRIME DESCRIPTION]
        \end{minipage}
    }
\end{center}

Users were asked: ``Do you think this defendant will commit a crime within 2 years?'', to which they could respond ``yes'' or ``no''. Users then received two forms of feedback: an indication showing whether their response was correct for that defendant, and a running total showing their accuracy over all defendants evaluated so far. 

Three catch trials were used to filter out users who were not paying close attention. The questions were formatted to look like the other questions but had easily identifiable answers. For example, ``The state of California was the 31st state to join the Union. California's nickname is: The Golden State ... Does the state of California have a nickname?''. If a user answered any of the catch-trials wrong, their data was removed from the dataset.

\subsection{Replication Study Results}
For our replication study, we repeated the no race experimental condition of Dressel and Farid's study, keeping the same set of defendants, the same catch-trials, and the same experimental protocol. 

We report results in a slightly different format than Dressel and Farid. The numbers reported in their paper \cite{dressel2018accuracy} are \emph{defendant-level}, where twenty user judgements for each defendant were aggregated into a score using majority voting aggregation. They then computed performance measures such as accuracy, false positive rates, and false negative rates at the defendant level.

In our work we do not aggregate different user judgements on the same defendant; rather, we report \emph{user-level} performance measures. This is core to our goal of ascertaining if a user's judgement of recidivism depends on how defendant race information is presented, and allows us to study how individual user characteristics and defendant characteristics may interact to impact user judgement. 

For Table \ref{tab:replication}, we recalculated \emph{user-level} performance measures on Dressel and Farid's data\footnote{Dressel and Farid's data is available at \url{www.cs.dartmouth.edu/farid/downloads/publications/scienceadvances17}.} using individual users' ``yes'' or ``no'' predictions of recidivism, and compare it to the defendant's ground truth recidivism outcome. We also calculated the same measures on our replication study data.

Comparing the results from our replication study to Dressel and Farid's results, Wilcoxon Signed-Ranks Tests revealed no significant differences for any of the measures (ACC: \textit{Z} = -1.08, \textit{p} = .281; FPR: \textit{Z} = -1.03, \textit{p} = .302; FNR:\textit{ Z} = -.392, \textit{p} = .695). Given that we successfully replicated Dressel and Farid's study, we move on to extend their work.

\begin{table}[!htbp]
  \resizebox{\linewidth}{!}{
  \begin{tabular}{l l c c}
    Metric & Defendant Race & Dressel \& Farid & Replication Study\\
    \midrule[1pt]
    \multirow{3}{*}{ACC} 
    &Overall &62.3\% &61.7\% \\
    &Black &62.2\% &62.2\% \\
    &White &63.2\% &61.6\% \\
    \midrule[1pt]
    \multirow{3}{*}{FPR}
    &Overall &39.8\% &40.3\% \\
    &Black &46.7\% &45.3\% \\
    &White &31.1\% &36.1\% \\
    \midrule[1pt]
    \multirow{3}{*}{FNR}
    &Overall &35.4\% &36.1\% \\
    &Black &31.1\% &32.2\% \\
    &White &42.7\% &42.3\% \\
  \end{tabular}}
  \caption{\textbf{Accuracy (ACC),  false positive rates (FPR), and false negative rates (FNR) from our replication study compared to Dressel and Farid's study, for the No Race condition. No significant differences were found between Dressel and Farid's study and our results.}}
  \label{tab:replication}
\end{table}

\section{Study 1: Do Race Labels and Photos Affect Judgements of Recidivism?}
In this study, we examined whether the form in which race information is presented to a user influences their judgements of recidivism. In the real world, judges have access to additional information outside of the defendant's file. The judge may see the defendant, hear them speaking, or even ask them questions.  Adding  photos to defendant profiles may be more realistic than just providing textual data and could influence users' decisions on whether or not a defendant will recidivate.

\subsection{Study 1 Experimental Protocol}
We designed a between-subjects study varying the presentation of race along two dimensions: Race Label {(on/off)} and Photo {(on/off)}. This resulted in four experimental conditions: \emph{No Race; Label Only; Photo Only; Label + Photo}. In their work, Dressel and Farid varied the Race Label (whether or not the defendant's race was listed in the text of their profile); this study extends their work by also varying whether or not a photo is displayed in the defendant's profile. Our study utilizes Dressel and Farid's experimental protocol (as described earlier) with a few minor modifications described below. Our study also underwent ethics, privacy, and IRB review. 

\subsubsection{Adding Photos} 
\begin{figure}[!htbp]
  \centering
  \includegraphics[width=0.7\columnwidth]{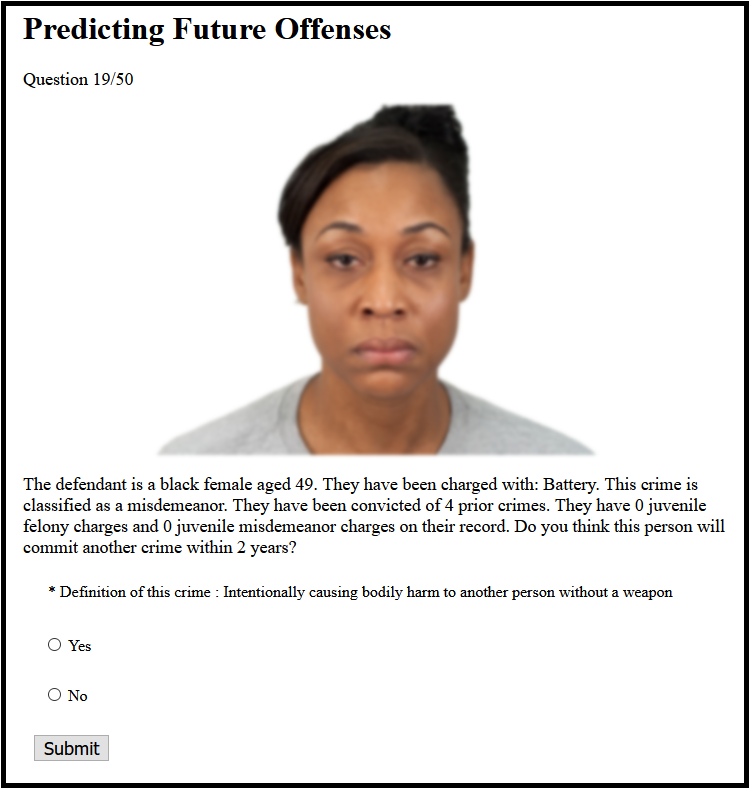}
  \caption{\textbf{Screenshot of the Label + Photo condition in Study 1. Users read the defendant's profile and were asked to predict if the defendant will commit a crime within two years.}}
  \label{fig:screenshot1}
\end{figure}

Defendant race labels are recorded in the ProPublica dataset \cite{propublica}, however photos of the actual defendants are not available. We therefore utilized images from a publicly available dataset: the Chicago Face Database (CFD) \cite{ma2015chicago}, which has been designed for use in scientific research. We matched each defendant with a photo corresponding to their age, race, and gender. Although the photo was not of the real defendant, the benefit of using these photos was that they were all captured in a consistent manner, and have similar visual characteristics. The CFD database was large enough that we were able to find representative photos for most of the defendants in our dataset. Some photos were reused for different defendants but we ensured that no user saw the same photo twice. We also applied a 1.75px blur to the photos so they looked less professional. Figure \ref{fig:screenshot1} shows a screenshot of a defendant profile. Users were not aware that the photos were not of the actual defendants until the end of the study when it was disclosed during a post-study debrief. 

\subsubsection{Feedback and Payment Modifications} 
We made two modifications to Dressel and Farid's experimental protocol. First, we removed feedback telling users whether they were correct or not after evaluating each defendant, and we omitted the running accuracy score. These modifications were made because we were concerned that feedback would cause variation in user performance across the session \cite{mason2012conducting,holsinger2018rejoinder}. Additionally, this modification addresses Dressel and Farid's concern that users may lose motivation during the task if they feel they are not going to achieve a monetary bonus (given to users who exceed a certain overall accuracy). 

Our second modification was increasing the payment to users to be \$1.50 (from \$1.00 paid by Dressel and Farid) for completion of the task, with a \$5.00 bonus (same as Dressel and Farid) for reaching an overall accuracy of 65\% or higher. This change was made to ensure that we were paying ethical wages \cite{mason2012conducting}. On average, the task was expected to take 10 minutes to complete, so workers would earn \$9 or \$39 per hour\footnote{If bonus is received, the estimated hourly rate is \$39=\$(1.50+5)*6.}, depending on whether they received the bonus. 

\subsubsection{Setup} 
The data for Study 1 was collected from 1,600 Amazon Mechanical Turk workers, four hundred per experimental condition. We used the same 1,000 defendants as Dressel and Farid, grouped into twenty sets of fifty defendants. Each set of defendants was rated by twenty users, for a total of four hundred users per condition (\textit{No Race, Label Only, Photo Only, Label+Photo}). 

Users ranged in age from 18 to over 74, with roughly even numbers of users who identified as females and males (50.7\% females; 49.2\% males). The majority of users self-identified as White (76.5\%), followed by Black (7.3\%), Asian (7.0\%), and Hispanic (6.2\%). Users also reported their education level, with 49.4\% holding a Bachelor's degree or higher 
and 46.25\% with education level below Bachelor's degree (with 4.35\% choosing not to answer). An additional 236 users completed the task but failed one or more catch trials and were removed from the dataset. 

\subsubsection{Analyses} 
Since our primary goal was to explore whether the presentation of race impacted users' judgements regarding recidivism, our main dependent variable was whether the defendants were predicted to recidivate. We also calculated accuracy, false positive rates, and false negative rates. Non-parametric statistics were used to analyse our data given that the data was not continuous and violated assumptions of homogeneity and sphericity. Non-parametric Kruskall-Wallis H one-way ANOVA tests were used to examine differences in recidivism predictions across the experimental conditions (\textit{No Race, Label Only, Photo Only, Label + Photo}) ($\alpha = .05$). Significant differences were followed-up using post-hoc pairwise Mann-Whitney U tests with Bonferroni corrections ($\alpha_{adjusted}= .008$).\footnote{The adjusted alpha is .008=0.05/6, since there are six pairwise comparisons across four experimental conditions.} Finally, to explore differences in recidivism predictions by group (e.g. user gender, defendant race, or defendant gender) or metric (e.g. accuracy, false positive rates), non-parametric Kruskall-Wallis H one-way ANOVA tests were applied within each group.

\subsection{Study 1 Results}
The percentage of defendants predicted to recidivate for each of the experimental conditions is shown in Table \ref{tab:study1all}. A statistically significant difference was found across the four conditions (\textit{H}(3) = 20.93, \textit{p} < 0.001). Post-hoc pairwise comparisons revealed that users were significantly less likely to predict that defendants would recidivate when both race labels and photos were shown, as compared to the other conditions (\textit{p} < 0.008). None of the other pairwise comparisons were statistically significant. 

\vspace{6pt}
\begin{mdframed}[nobreak=true, style=MyFrame]
\textbf{Result 1a:} Adding both photos and race labels to defendant profiles made users \textbf{less} likely to predict that a defendant would recidivate within two years.
\end{mdframed}

\begin{table}[!htbp]
  \resizebox{\linewidth}{!}{
  \begin{tabular}{cccc}
    No Race&Label Only&Photo Only& Label+Photo\\
    \midrule[1pt]
    \textbf{54.7}\%&\textbf{55.1}\%&\textbf{54.8}\%&\textbf{53.1\%*}\\
    \midrule[1pt]
    {*}\textit{p} < .008 
  \end{tabular}}
  \caption{\textbf{Study 1: Percentage of defendants predicted to recidivate across the four experimental conditions. Significant differences across conditions were found, with fewer defendants predicted to recidivate in the Label+Photo condition than the other three conditions.}}
  \label{tab:study1all}
\end{table}
 
\subsubsection{User Gender}
Prior research has shown that the gender of an Amazon Mechanical Turk worker can have a significant effect on results, particularly when gender or race is involved in the task \cite{jahanbakhsh2020bias, peng2019you}. We therefore investigated whether users' self-identified gender impacted recidivism predictions for our experimental conditions. Table \ref{tab:study1turkgender} shows the percentage of defendants predicted to recidivate in each condition, split by the gender of the user evaluating that defendant.

No significant differences across experimental conditions were found when the predictions were done by male users  (\textit{H}(3) = 3.03, \textit{p} = .388). However, for female users, predictions made across different experimental conditions were significantly different (\textit{H}(3) = 21.78, \textit{p} < 0.001). Post-hoc pairwise comparisons revealed that female users were less likely to predict that a  defendant would recidivate in the Label+Photo experimental condition than in the other three conditions (\textit{p} < .008). 

\vspace{6pt}
\begin{mdframed}[nobreak=true,style=MyFrame]
\textbf{Result 1b:} Adding both photos and race labels made female users \textbf{less} likely to predict that a defendant would recidivate within two years but had no significant effect on male users' predictions. 
\end{mdframed}
\vspace{6pt}

\begin{table}[!htbp]
  \resizebox{\linewidth}{!}{
  \begin{tabular}{lcccc}
    User\\
    Gender&No Race&Label Only&Photo Only& Label+Photo\\
    \midrule[1pt]
    Male&52.4\%&53.0\%&53.3\%&52.2\%\\
    \textbf{Female}&\textbf{56.6\%}&\textbf{57.0\%}&\textbf{56.5\%}&\textbf{54.0\%*}\\
    \midrule[1pt]
    {*}\textit{p} < .008 
    \end{tabular}}
  \caption{\textbf{Study 1: Percentage of defendants predicted to recidivate across the four experimental conditions, split by user-specified gender. Significant differences across conditions were found for female users, with fewer defendants predicted to recidivate in the Label+Photo condition than the other three conditions.}}
  \label{tab:study1turkgender}
\end{table}
 
Since no significant differences across experimental conditions were found for male users, 
subsequent analyses for Study 1 will focus on predictions made by female users. 

\subsubsection{Defendant Race}
We explored whether the defendant's race had an impact on recidivism predictions. Given that there were very few Hispanic, Asian and Native American defendants in the dataset, profiles belonging to these race categories were grouped together, resulting in three race categories: White (n=377), Black (n=530) and Other (n=93).

The percentage of defendants predicted by female users to recidivate in each experimental condition, by race, is shown in Table \ref{tab:study1defendantrace}. No significant differences across the experimental conditions were found for white defendants ($H(3) = 2.81, p = .422$) and defendants of other races ($H(3) = 3.67, p = .300$), but a significant difference was found for black defendants ($H(3) = 23.21, p < 0.001$). Post-hoc pairwise comparisons revealed that female users were less likely to predict that a black defendant would recidivate in the Label + Photo condition than in the other three conditions ($p < .008$).     

\vspace{6pt}
\begin{mdframed}[nobreak=true,style=MyFrame]
\textbf{Result 1c:} Adding both photos and race labels only impacted black defendants, and not white defendants or defendants of other races. The photos and race labels 
made female users 
less likely to predict that a black defendant would recidivate within two years. 
\end{mdframed}
\vspace{6pt}

\begin{table}[!htbp]
  \resizebox{\linewidth}{!}{
  \begin{tabular}{lcccc}
    Defendant\\
    Race&No Race&Label Only&Photo Only& Label+Photo\\
    \midrule[1pt]
    White&49.3\%&50.6\%&49.6\%&48.8\%\\
    \textbf{Black}&\textbf{63.3\%}&\textbf{62.9\%}&\textbf{62.7\%}&\textbf{59.2\%*}\\
    Other&48.0\%&49.2\%&48.5\%&45.0\%\\
    \midrule[1pt]
  {*}\textit{p} < .008 
  \end{tabular}}
  \caption{\textbf{Study 1: Percentage of defendants predicted by female users to recidivate across the four experimental conditions, split by defendant race. The "Other" race category included Hispanic, Asian, and Native American defendants. Significant differences across conditions were found for black defendants, with fewer defendants predicted to recidivate in the Label+Photo condition than the other three conditions.}}
  \label{tab:study1defendantrace}
\end{table}

\subsubsection{Defendant Gender}
\label{sec:defendant_gender}
We explored whether the gender of the defendant (as specified in the ProPublica dataset \cite{propublica}) impacted recidivism predictions made by users who identified as female, across experimental conditions. Note however that there were fewer female defendants (n=197) in our dataset than male defendants (n=803), so our statistical power for detecting differences for female defendants was lower. The percentage of defendants predicted by female users to recidivate in each experimental condition, split by defendant gender, are shown in Table \ref{tab:study1defendantgender}. A significant difference in experimental condition was found for male defendants ($H(3) = 21.62, p <.001$), but not female defendants ($H(3) = 2.32, p = .509$). Post-hoc pairwise comparisons revealed that female users were less likely to predict that a male defendant would recidivate in the Label + Photo condition than in the other conditions ($p < .008$).     

\vspace{6pt}
\begin{mdframed}[nobreak=true, style=MyFrame]
\textbf{Result 1d:} Adding both photos and race labels impacted male defendants, making female users less likely to predict that a male defendant would recidivate within two years.
\end{mdframed}
\vspace{6pt}

\begin{table}[!htbp]
  \resizebox{\linewidth}{!}{
  \begin{tabular}{lcccc}
    Defendant\\
    Gender&No Race&Label Only&Photo Only& Label+Photo\\
    \midrule[1pt]
    \textbf{Male}&\textbf{59.7\%}&\textbf{59.7\%}&\textbf{59.4}\%&\textbf{56.6\%*}\\
    Female&43.8\%&45.8\%&44.7\%&43.7\%\\
    \midrule[1pt]
    {*}\textit{p} < .008
  \end{tabular}}
  \caption{\textbf{Study 1: Percentage of defendants predicted by female users to recidivate across the four experimental conditions, split by defendant gender. Significant differences across conditions were found for male defendants, with again fewer defendants predicted to recidivate in the Label+Photo condition than the other three conditions.}}
  \label{tab:study1defendantgender}
\end{table}

\subsubsection{Accuracy, False Positives and False Negatives}
Finally, we examined whether or not the different experimental conditions impacted accuracy, false positive or false negative rates for all users (see Table \ref{tab:study1accuracy}). There was no significant difference in accuracy across the conditions ($H(3) = 3.218, p =.359$), however there were significant differences in the false positive rates ($H(3) = 14.27, p =.003$) and false negative rates ($H(3) = 11.01, p = .012$) with the Label + Photo condition having a lower false positive rate, and a higher false negative rate. As expected, further analyses revealed that these differences were only observed for black defendants, and not for white defendants or defendants of other races.

\vspace{6pt}
\begin{mdframed}[nobreak=true,style=MyFrame]
\textbf{Result 1e:} Adding both photos and race labels had no impact on accuracy, but lowered false positive rates and increased false negative rates for black defendants.
\end{mdframed}
\vspace{6pt}

\begin{table}[!htbp]
  \resizebox{\linewidth}{!}{
  \begin{tabular}{llcccc}
    &Defendant\\
    Metric&Race&No Race&Label Only&Photo Only& Label+Photo\\
    \midrule[1pt]
    \multirow{3}{*}{ACC}& Overall &62.5\%&62.6\%&61.9\%&62.6\%\\
    & Black &63.1\%&63.9\%&63.3\%&63.7\%\\
    & White &62.4\%&61.6\%&60.5\%&61.6\%\\
    \midrule[1pt]
    \multirow{3}{*}{\textbf{FPR}}& Overall &42.5\%&42.9\%&43.2\%&40.9\%\\
     & \textbf{Black}&\textbf{48.6\%}&\textbf{47.7\%}&\textbf{47.3\%}&\textbf{44.7\%*}\\
     & White&37.8\%&38.7\%&40.0\%&38.2\%\\
    \midrule[1pt]
    \multirow{3}{*}{\textbf{FNR}}& Overall&32.0\%&31.4\%&32.5\%&33.5\%\\
     &\textbf{Black}&\textbf{28.3\%}&\textbf{27.5\%}&\textbf{28.7\%}&\textbf{30.0\%*}\\
     &White&37.4\%&37.8\%&38.8\%&38.8\%\\
    \midrule[1pt]
    {*}\textit{p} < .008
  \end{tabular}}
  \caption{\textbf{Study 1: Accuracy (ACC), false positive rates (FPR), and false negative rates (FNR) across the four experimental conditions, split by defendant race. Significant differences were found for black defendants with lower false positive rates and higher false negative rates in the Label+Photo condition than the other three conditions.}}
  \label{tab:study1accuracy}
 \end{table}

\subsection{Study 1 Discussion}
Showing the race of a defendant using both a text label and a photo significantly impacted female users' assessments of black defendants, but had no impact on white or other race defendants. No significant differences in assessments of defendants were found for users who identified as male. Given that previous literature has noted bias against black defendants for recidivism~\cite{propublica, skeem2016risk, chouldechova2017fair}, we expected that racial cues would have \emph{increased} predictions of recidivism, however, our results were the opposite. Users were \emph{less likely} to predict that a black defendant would recidivate within two years when a race label and photo were included in the defendant description. 

Interestingly, neither the race label nor the photo on their own impacted the recidivism predictions. For the race label, we hypothesize that the saliency of this features is low. It is a small word in a block of text; it could be easily be overlooked by the user. While the photo is a more salient signal, it didn't make a difference on its own. We hypothesize that it can be difficult to infer race from a photo alone. As such, it was the combination of both the race label and the photo that had a significant effect on recidivism predictions. 

Our results revealed significant differences for male defendants but not female defendants, however, the number of female defendants in our dataset was low, and we did not systematically assign user gender to defendants. Our next study further explores the issue of defendant gender. 

Examining the recidivism predictions in Table \ref{tab:study1defendantrace}, we see that black defendants had an overall higher prediction of recidivism compared to the other races. This may be related to the defendant profiles, since ground truth recidivism was also higher among black defendants in the dataset compared to defendants of other races in the dataset -- 57\% of black defendants were reported to have recidivated compared to 37\% of white defendants and 37\% of defendants of other races. Hence, it is possible that the differences we observed after adding the race label and photos to defendant profiles were related to other defendant features rather than race (i.e. certain defendant profiles have higher ground truth recidivism rates). Hence, we designed our next study to isolate the effect of defendant race.

\section{Study 2: Isolating Effect of Defendant Race} 
In this study we modified the experimental design from Study 1 to isolate the effect of defendant race and continue our investigation into whether additional racial information influences users’ judgements of recidivism. In particular, we wanted to understand if adding race labels and photos has a differential impact on black versus white defendants, \textit{independent of non-race defendant features}. Additionally, we increased the female defendants in our dataset to better understand if the impact of adding race labels and photos is different for male and female defendants. 

\subsection{Study 2 Experimental Protocol}
We designed a within-subjects study where we controlled the race for each defendant profile, holding all other features constant. Each defendant profile was evaluated multiple times, once as a black defendant, once as a white defendant, and once with no race information provided. We then examined differences in the recidivism predictions across these three experimental conditions (\textit{No Race, White Label+Photo, Black Label+Photo}). This important modification helped ensure that any differences observed were a result of the racial representation (the label and photo) and not related to characteristics of the defendant profile itself. We provide more details on our experimental design below. 

\subsubsection{Assignment of Defendant Race}
\begin{figure}[!htbp]
  \includegraphics[width=0.35\textwidth]{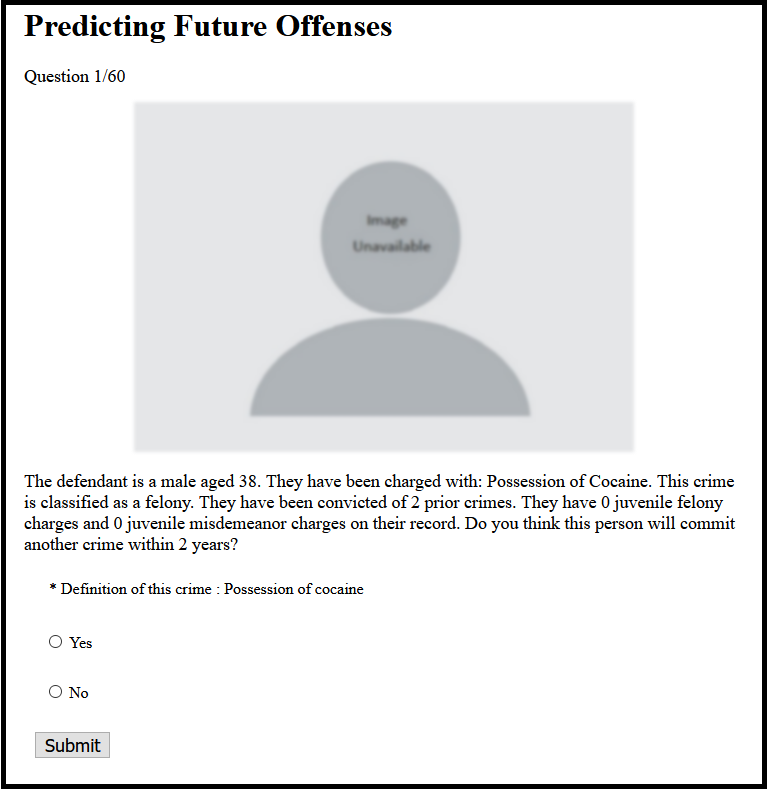}
   \centering
  \caption{\textbf{Screenshot of the No Race condition in Study 2. Users read the defendant's profile and were asked to predict if the defendant will commit a crime within two years.}}
    \label{fig:screenshot2}
  
\end{figure}

Each experimental session involved sixty defendant profiles -- twenty unique defendant profiles (ten male and ten female), presented in three trial clusters. In the first trial cluster, each defendant was randomly assigned to be either a black defendant, a white defendant, or have no race information. In the second trial cluster, one of the remaining representations was randomly chosen for each profile, and in the third cluster the final representation was chosen. The result of this within-subjects design was a presentation of sixty defendant profiles to each user, where the order was mixed in terms of the race of the defendant (Black, White, or No Race). For the no race instances, the race label was removed from the profile text and a blank photo icon was shown which stated ``Image Unavailable'' instead of a photo (see Figure \ref{fig:screenshot2}).

There was no time gap or delineating features between the different trial clusters, to give the appearance of sixty unique defendants. Given the similarities across profiles, and the length of the task, many users likely did not realize that the same profile was being presented multiple times. If however some users did recognize the profiles, this could reduce the likelihood of detecting differences between the conditions.

\subsubsection{Balancing Defendant Gender} As pointed out before, Dressel and Farid's sample of 1,000 defendants has fewer female defendants (n=197) than male defendants (n=803), which reduces the statistical power for detecting differences for the smaller group of female defendants. To ensure that we had equal numbers of male and female defendants in this study, we selected additional defendants from the original ProPublica COMPAS dataset \cite{propublica}, resulting in eight hundred male and eight hundred female defendant profiles, split evenly across black and white defendants. 

\subsubsection{Setup}
The data for Study 2 was collected from 1,600 Amazon Mechanical Turk workers 
(eight hundred who self-identified as female and eight hundred who self-identified as male), with equal numbers of females and males evaluating each defendant. 

Similar to Study 1, users ranged in age from 18 to over 74. The majority of users self-identified as White (71.6\%), followed by Black (10.2\%), Asian (8.7\%), and Hispanic (5.4\%). Users were also asked to report their education level, with 54.4\% holding a Bachelor's degree or higher 
and 45.0\% with education level below Bachelor's Degree (and 0.6\% choosing not to answer). An additional 635 users completed the task but failed at least one of the catch trials and were removed from the dataset. 

\subsubsection{Analyses}
Similar to Study 1, our primary dependant measure was whether or not defendants were predicted to recidivate. We also analyzed accuracy, false positive rates, and false negative rates. Non-parametric statistics were used to analyse our data given that the it was not continuous and violated assumptions of homogeneity and sphericity. To account for the within-subjects design, non-parametric Friedman Tests were used to examine differences in recidivism predictions across the experimental conditions (\textit{No Race, White Label+Photo, Black Label+Photo}) ($\alpha = .05$). Significant differences were followed-up using posthoc pairwise Wilcoxon Signed Rank Tests with Bonferroni corrections ($\alpha_{adjusted}= .017$).\footnote{The adjusted alpha is .0017=0.05/3, since there are three pairwise comparisons across three experimental conditions.} Finally, to study if the difference in recidivism predictions across the three experimental conditions differed by group (e.g. user gender, defendant race, or defendant gender) or metric (e.g. accuracy, false positive rates), non-parametric Friedman tests were also applied, but within each group. 

\subsection{Study 2 Results}
The percentage of defendants predicted to recidivate for each condition is shown in Table \ref{tab:study2all}. Our results revealed a significant differences across experimental conditions (${\chi}^2 (2) = 43.95, p < 0.001$). All post-hoc pairwise comparisons were significant ($p < .017$). Similar to Study 1, when a race label and a photo were shown to users, they were significantly less likely to predict that a black defendant would recidivate compared to a white defendant, even though the profiles were identical. 
In contrast to the results from Study 1, users were significantly less likely to predict that a defendant would recidivate when no race information was provided ($p < .017$). 

\vspace{6pt}
\begin{mdframed}[nobreak=true,style=MyFrame]
\textbf{Result 2a:} Even when users rated the same defendant profile, they were \textbf{less} likely to predict that a defendant would recidivate when the defendant's race was presented as black instead of white. 
\end{mdframed}
\vspace{6pt}

\begin{table}[!htbp]
  \resizebox{\linewidth}{!}{
  \begin{tabular}{ccc}
    No Race &White Label+Photo& Black Label+Photo\\
    \midrule[1pt]
    \textbf{53.4\%*}&\textbf{55.0\%*}&\textbf{54.0\%*}\\
    \midrule[1pt]
    {*}\textit{p} < .017
  \end{tabular}}
  \caption{\textbf{Study 2: Percentage of defendants predicted to recidivate across the three experimental conditions. All pairwise comparisons were significant with fewer defendants predicted to recidivate in the Black Label+Photo condition than the White Label+Photo condition, and even fewer defendants predicted to recidivate in the No Race condition. }}
  \label{tab:study2all}
\end{table}

\subsubsection{User Gender}
We again examined whether the self-identified gender of the user impacted recidivism predictions. The percentage of defendants predicted to recidivate in each condition, by user gender, are shown in Table \ref{tab:study2turkgender}. Similar to Study 1, there were no significant difference across conditions for male users (${\chi}^2 (2) = 4.81, p = 0.090$). However, for female users, the percentage of defendants predicted to recidivate differed significantly across conditions (${\chi}^2 (2) = 53.51, p < 0.001$). All post-hoc pairwise comparisons were significant ($p < .017$) with the No Race condition having the fewest predictions of recidivism. Similar to Study 1, the presented race of the defendant mattered to female users and they were less likely to predict that a defendant would recidivate when they were told it was a black defendant, compared to a white defendant. 

\vspace{6pt}
\begin{mdframed}[nobreak=true,style=MyFrame]
\textbf{Result 2b:} Presenting a defendant's race as black instead of white made 
female users less likely to predict that the defendant would recidivate, but had no significant effect on 
male users' predictions. 
\end{mdframed}
\vspace{6pt}

\begin{table}[!htbp]
  \resizebox{\linewidth}{!}{
  \begin{tabular}{lccc}
    User\\
    Gender&No Race& White Label+Photo&Black Label+Photo\\
    \midrule[1pt]
    Male&52.2\%&52.9\%&52.4\%\\
    \textbf{Female}&\textbf{54.8\%*}&\textbf{57.0\%*}&\textbf{55.7\%*}\\
    \midrule[1pt]
    {*}\textit{p} < .017
  \end{tabular}}
  \caption{\textbf{Study 2: Percentage of defendants predicted to recidivate across the three experimental conditions, split by user-specified gender. Significant differences across conditions were found for female users, with fewer defendants predicted to recidivate in the Black Label+Photo condition than the White Label+Photo condition. All pairwise comparisons were significant.}}
  \label{tab:study2turkgender}
\end{table}

\subsubsection{Defendant Gender}
\label{sec:defendant_gender}
We next explored whether the gender of the defendant (as specified in the ProPublica dataset \cite{propublica}) impacted recidivism predictions. The percentage of female and male defendants predicted to recidivate by female users are shown in Table \ref{tab:study2defendantgender}. Significant differences across conditions were found for both male and female defendants (${\chi}^2 (2) = 27.81, p < 0.001$, and (${\chi}^2 (2) = 30.77, p < 0.001$, respectively). Post-hoc pairwise comparisons revealed that for male defendants, there was no difference when they were portrayed as being white or black, but they were less likely to be predicted to recidivate by female users when no race information was provided. For female defendants, they were less likely to be predicted to recidivate by female users if they were portrayed as being black, or with no race information ($p < .017$).

\vspace{6pt}
\begin{mdframed}[nobreak=true, style=MyFrame]
\textbf{Result 2c:} Adding photos and race labels impacted both male and female defendants. For male defendants, the additional race information made 
female users 
more likely to predict that they would recidivate, irrespective of the portrayed race. For female defendants, they were more likely to be predicted to recidivate when they were portrayed as being white, compared to black, or when no race information was provided.
\end{mdframed}
\vspace{6pt}

\begin{table}[!htbp]
  \resizebox{\linewidth}{!}{
  \begin{tabular}{lccc}
    Defendant\\
    Gender&No Race& White Label+Photo&Black Label+Photo\\
    \midrule[1pt]
    \textbf{Male}&\textbf{60.8\%*}&\textbf{63.1\%}&\textbf{62.4\%}\\
    \textbf{Female}&\textbf{48.7\%}&\textbf{51.0\%*}&\textbf{49.0\%}\\
    \midrule[1pt]
    {*}\textit{p} < .017
  \end{tabular}}
  \caption{\textbf{Study 2: Percentage of defendants predicted by female users to recidivate across the three experimental conditions, split by defendant gender. Significant differences across conditions were found regardless of defendant reported gender; statistically significant pairwise comparisons are described above.}}
  \label{tab:study2defendantgender}
\end{table}

\subsubsection{Accuracy, False Positives and False Negatives}
We again examined whether or not the experimental conditions impacted accuracy, false positive and false negative rates for all users (see Table \ref{tab:study2accuracy}). Significant differences were found for all metrics (ACC: ${\chi}^2 (2) = 10.16, p = 0.006$; FPR: ${\chi}^2 (2) = 53.51, p < 0.001$; and FNR: ${\chi}^2 (2) = 7.50, p = 0.023$). Post-hoc pairwise analyses revealed that when photos and race labels were provided, users had higher accuracy, lower false positive rates, and higher false negative rates for black defendants, compared to white defendants. Results for the no race condition were similar to the black defendants except the false positive rate was significantly lower. 

\vspace{6pt}
\begin{mdframed}[nobreak=true,style=MyFrame]
\textbf{Result 2d:}  Presenting a defendant's race as black instead of white increased accuracy, decreased false positive rates, and increased false negative rates. 
\end{mdframed}
\vspace{6pt}

\begin{table}[!htbp]
  \resizebox{\linewidth}{!}{
  \begin{tabular}{llccc}
    Metric & No Race & White Label+Photo& Black Label+Photo\\
    \midrule[1pt]
    \multirow{1}{*}{\textbf{ACC}} & \textbf{60.8\%} & \textbf{60.1\%*} & \textbf{60.4\%} \\
    \multirow{1}{*}{\textbf{FPR}} & \textbf{43.3\%*}&\textbf{45.3\%*}&\textbf{44.2\%*}\\
    \multirow{1}{*}{\textbf{FNR}} & \textbf{34.2\%}&\textbf{33.3\%*}&\textbf{34.0\%}\\
    \midrule[1pt]
    {*}\textit{p} < .017
  \end{tabular}}
  \caption{\textbf{Study 2: Accuracy, false positive rates, and false negative rates across the three experimental conditions. Significant differences across conditions were found for all metrics. All pairwise comparisons were significant for the false positive rate metric. For accuracy and false negative rates, the White Label+Photo condition was significantly lower than the other conditions.}}
  \label{tab:study2accuracy}
\end{table}

\subsection{Study 2 Discussion}
Similar to Study 1, when race labels and photos were added to defendant profiles, users were less likely to predict that 
black defendants would recidivate compared white defendants. 
This provides evidence that the observed differences are attributable to the presented race of the defendant, and not to other features inherent in the profiles. Additionally, Study 2 replicated the result that 
users who identified as female were sensitive to the race of the defendant, but not those who identified as male. The added female defendants in our dataset showed that these differences are applicable to both male and female defendants (not just males as suggested in Study 1). 

One difference in the results between Study 1 and Study 2 was related to the No Race condition. In Study 1, users were \textbf{more} likely to predict that defendants in the No Race condition would recidivate, compared to the Label+Photo condition; however, in Study 2, users were \textbf{less} likely to predict that defendants in the No Race condition would recidivate, compared to the other conditions. There were two differences which may have contributed to this change. First, users in Study 1 were only exposed to one condition, while users in Study 2 were exposed to all three conditions, and the trials were intermixed. Second, the visual layout of the tasks were different in the two studies. In Study 1, the No Race condition only showed a block of text describing the defendant's profile. In Study 2 a placeholder ``Image Unavailable'' photo was added (see Figure 2). We hypothesize that if photos have a humanizing effect on the task (as suggested in Study 1), having trials with photos adjacent to the No Race trials may cause a ``spillover'' humanizing affect, making users more conscious about the people they were evaluating in the task. Alternatively, the ``Image Unavailable'' icon may have altered users perceptions of those defendants. 

In Study 1 users were presented with fifty unique defendants, but in Study 2 users were presented with twenty unique defendants that they evaluated three times. We examined the level of agreement users had across the conditions. While a reasonable level of agreement is expected if users were being consistent in their predictions, we would expect some variability given the complexity of the task. Overall, only 0.3\% of users (n=50) had perfect agreement across their predictions, and on average, the level of agreement was 75\%. If users did notice the repetition of defendants, they would be more likely to repeat their predictions, making it more difficult to detect differences between conditions. Given that we still observed significant differences provides support for these results.

\section{Results Summary}
Although we hypothesized that adding defendant photos would impact users' judgements, we were surprised in the ways it manifested in our results.

\subsection{Adding Photos Mattered}
In contrast to Dressel and Farid’s findings \cite{dressel2018accuracy}, showing race information had a significant impact on users’ recidivism predictions but only when both a photo and race label were added to defendants' profiles. It is interesting that neither the label nor the photo on their own made a difference, but the combination of the two had a significant impact on users' judgement. 

We initially hypothesized that adding photos might induce racial bias and therefore would \textbf{increase} recidivism predictions for certain classes of defendants. However, in Study 1, adding labels and photos \textbf{reduced} recidivism predictions. This suggests that photos may have a humanizing effect on the task, making users less likely to predict that a defendant will recidivate. 

\subsection{No Race Conditions}
It is unclear what caused the No Race condition to be rated higher (than the race conditions) in one study, and lower (than the race conditions) in another study. As previously explained, there were small differences in this condition between the two studies which may have impacted the results. Regardless, whether it was the visual presentation of the task (i.e. the ``Image Unavailable'' icon), or the fact that the race trials were intermixed, it suggests that human judgements are susceptible to even small design choices. Further research is needed to better understand the difference in these results.

\subsection{Female vs Male Workers on Amazon Mechanical Turk}
One of the strongest results from this work are the striking differences between 
users who identified as female versus male. In both studies, significant differences across experimental conditions were observed for female users but not males. This is consistent with other recent studies \cite{peng2019you, jahanbakhsh2020bias}, where female and male Mechanical Turk Workers reacted differently to issues related to race and gender. This is an important takeaway for researchers who use Amazon Mechanical Turk. 

\subsection{Differential Impact of Race}
The results for black defendants were particularly interesting, especially given that there is well documented bias against black people in the criminal justice field~\cite{propublica, skeem2016risk, smith1984equity, schwalbe2006classifying, rachlinski2008does}, and other research on recidivism that suggests that users rate black defendants more harshly~\cite{kleider2012looking, kleider2017black, dixon2005skin, green2019disparate}. The opposite was true in our studies, as the addition of race labels and photos reduced the likelihood that a black defendant would be predicted to recidivate. We hypothesized that users may have overcompensated to ensure they were not exhibiting a common societal bias.  

\subsection{Practical Significance}
Many of the significant results uncovered in these studies only represent a small difference in the actual population percentages. Although this suggests that the results have low practical significance, when it comes to issue of racial and gender bias, making things better for even a small number of people is important. As such, we value any improvement we can make, even if the scope is small.

\subsection{Study Limitations}

\subsubsection{Recidivism Predictions as a Mechanical Turk Task} 
We are not proposing that recidivism decisions should be done by Amazon Mechanical Turk workers. Recidivism is a complex criminal justice process that requires expert knowledge. However, in the context of this study, having Mechanical Turk workers assess the likelihood of someone committing a crime in the future is a reasonable lay-task. Juries are made up of everyday people who are asked to assess guilt or innocence. Also, as people go about their daily lives they often need to assess whether the people around them are law abiding citizens, or potential criminals. Previous work by Dressel and Farid provides an experimental framework to explore issues around bias that allows for different controls and extensions. 

\subsubsection{Defendant Photos}
Our use of photos from the Chicago Face Database is both a benefit, and a limitation. Being able to use photos of real defendants would give us more accurate results; however, those photos could also be heavily biased as they would have been taken under widely varying conditions. Additionally, it is unethical to use photos from people who have not consented to their use. Getting access and consent to use photos from a large corpus of defendants would be difficult. In contrast, the CFD photos were collected with the intent to be used in scientific research.

\subsubsection{User Race Interactions}
Our results show interaction between race and gender, both for the defendants and for the users evaluating the defendants. There is a strong likelihood that users' race would also impact their judgments, and probably interact with the race and gender of the defendants. While we were able to collect a balanced sample of users who identified as female and male, approximately 75\% of our users self-identified as white. Collecting a racially diverse set of users is challenging but should be pursued in future work. 

\subsubsection{Ground Truth}
Although our results primarily focused on perceived criminality across conditions, we also reported on accuracy, false positive, and false negative rates. It is important to understand that these measures all compare human judgement to the recorded ground truth for the defendant; however, ground truth for recidivism is extremely noisy and heavily biased \cite{smith1984equity}. This raises concerns about the validity of ground truth recidivism data, and therefore measures of accuracy, false positive and false negative rates should be interpreted with caution.

\subsubsection{Binary Gender}
This study only reports results with male and female defendant profiles and Amazon Mechanical Turk workers who self-identified as male or female. While we recognize that there are many ways people self-identify their gender, the majority of our participants identified as either male or female, and the dataset used for defendant profiles only reported binary genders. We acknowledge the importance of studies that consider gender as a non-binary variable and hope that future studies will take this into consideration.

\section{Conclusions}
In conclusion, this work demonstrates that human judgement is impacted by how race information is presented, and sometimes in ways that contradict our expectations. By adding both race labels and photos to defendant profiles, users were made more aware of the defendant's race. However, instead of exacerbating racial biases, it made users \textbf{less} likely to view someone as a criminal.

This work also demonstrated an interaction between gender and race, as it was the female users in our study who were influenced by the additional race information. This resulted in the female users rating black defendants as less likely to recidivate compared to white defendants. Further research is needed to better understand why defendant race did not impact judgements of male users. 

The results of our work suggest that system-designers need to develop ways to account for and mitigate human biases in development of future online systems. Since there was differential impact found based on interactions between the system users and the individuals they are asked to evaluate, a potential solution could involve personalizing tasks to different types of users to account for these interactions. Moreover, biases not only manifest in humans but also in algorithms, so system-designers should also consider Human + AI systems that complement each other and maximize their potential. 

Overall, this paper extends the conversation started by Dressel and Farid and opens the door for more research investigating how representations influence human judgement. More understanding in this space is critical, especially given the growth of AI systems that need to work \textit{with} users to help inform societal decisions. 

\section{Acknowledgements}
We would like to thank Justin Cranshaw, Scott Counts, Gabrielle Quinn, and Alexa Brewer for their assistance with this project. We also want to thank the reviewers for their valuable feedback.

\balance{}

\bibliographystyle{SIGCHI-Reference-Format}
\bibliography{reference}

\end{document}